# Cheap Method for Shielding a City from Rocket and Nuclear Warhead Impacts*

**Alexander Bolonkin**
C&R, 1310 Avenue R, #F-6, Brooklyn, NY 11229, USA
T/F 718-339-4563, aBolonkin@gmail.com, http://Bolonkin.narod.ru

## Abstract

   The author suggests a cheap closed AB-Dome which protects the densely populated cities from nuclear, chemical, biological weapon (bombs) delivered by warheads, strategic missiles, rockets, and various incarnations of aviation technology. The offered AB-Dome is also very useful in peacetime because it shields a city from exterior weather and creates a fine climate within the AB-Dome. The hemispherical AB-Dome is the inflatable, thin transparent film, located at altitude up to as much as 15 km, which converts the city into a closed-loop system. The film may be armored the stones which destroy the rockets and nuclear warhead. AB-Dome protects the city in case the World nuclear war and total poisoning the Earth's atmosphere by radioactive fallout (gases and dust). Construction of the AB-Dome is easy; the enclosure's film is spread upon the ground, the air pump is turned on, and the cover rises to its planned altitude and supported by a small air over-pressure. The offered method is cheaper by thousand times than protection of city by current anti-rocket systems. The AB-Dome may be also used (height up to 15 and more kilometers) for TV, communication, telescope, long distance location, tourism, high placed windmills (energy), illumination and entertainments. The author developed theory of AB-Dome, made estimation, computation and computed a typical project. Discussion and results are in the end of article.

**\*** Presented to http://arxiv.org on Jan.10, 2008.

## Introduction
1. **Effects of nuclear explosion.**

The dominant effects of a nuclear weapon where people are likely to be directly affected (blast and thermal radiation) are identical physical damage mechanisms to conventional explosives. However the energy produced by a nuclear explosive is millions of times more powerful per gram and the temperatures reached are, briefly, Sun-like.  (See: G.I. Brown's THE BIG BANG: A HISTORY OF EXPLOSIVES, 1998.)

Energy from a nuclear explosive is initially released in several forms of penetrating radiation. When there is a surrounding material such as air, rock, or water, this radiation interacts with and rapidly heats it to an equilibrium temperature. This causes vaporization of surrounding material resulting in its rapid expansion. **Kinetic energy** created by this expansion contributes to the formation of a **shockwave**. When a nuclear detonation occurs in air near sea level, much of the released energy interacts with the atmosphere and creates a shockwave which expands spherically from the hypocenter. Intense **thermal radiation** at the hypocenter forms a fireball and if the burst is low enough, it is often associated an aerial mushroom cloud. In a burst at high altitudes, where the air density is low, more energy is released as ionizing gamma radiation and x-rays than an atmosphere displacing shockwave.

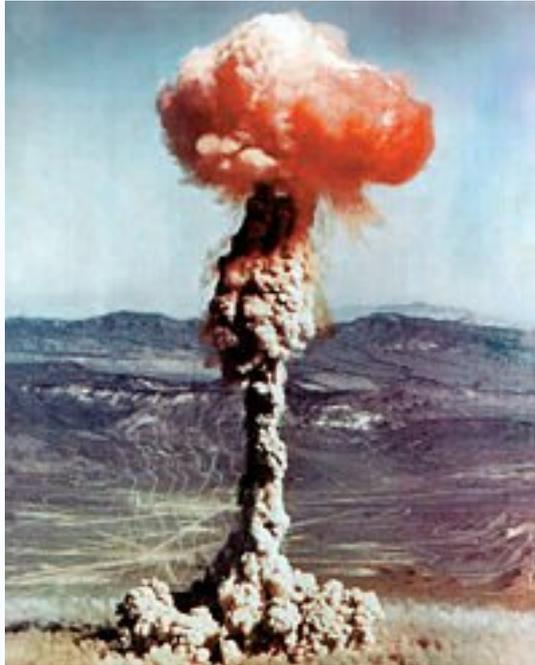
A low-yield USA fission nuclear test in Nevada

**Shockwave**. The high temperatures and pressures cause gas to move outward radially in a thin, dense shell called "the hydrodynamic front." The front acts like a piston that pushes against and compresses the surrounding medium to make a spherically expanding shock wave. At first, this shock wave is inside the surface of the developing fireball, which is created in a volume of air by the X-rays. However, within a fraction of a second the dense shock front obscures the fireball, making the characteristic double pulse of light seen from a nuclear detonation. For air bursts at or near sea-level between 50-60% of the explosion's energy goes into the blast wave, depending on the size and the yield-to-weight ratio of the bomb. As a general rule, the blast fraction is higher for low yield and/or high bomb mass. Furthermore, it decreases at high altitudes because there is less air mass to absorb radiation energy and convert it into blast. This effect is most important for altitudes above 30 km, corresponding to <1 per cent of sea-level air density.

Much of the destruction caused by a nuclear explosion is due to blast effects. Most buildings, except reinforced or blast-resistant structures, will suffer moderate to severe damage when subjected to overpressures of only 35.5 kilopascals (kPa).

The blast wind may exceed several hundred km/h. The range for blast effects increases with the explosive yield of the weapon and also depends on the burst altitude. Contrary to what one might expect from geometry the blast range is not maximal for surface or low altitude blasts but increases with altitude up to an "optimum burst altitude" and then decreases rapidly for higher altitudes. This is due to the nonlinear behaviour of shock waves. If the blast wave reaches the ground it is reflected. Below a certain reflection angle the reflected wave and the direct wave merge and form a reinforced horizontal wave, the so-called Mach stem (named after Ernst Mach). For each goal overpressure there is a certain optimum burst height at which the blast range is maximized. In a typical air burst, where the blast range is maximized for 35 to 140 kPa, these values of overpressure and wind velocity noted above will prevail at a range of 0.7 km for 1 kiloton (kt) of TNT yield; 3.2 km for 100 kt; and 15.0 km for 10 megatons (Mt) of TNT.

Two distinct, simultaneous phenomena are associated with the blast wave in air:

- **Static overpressure**, i.e., the sharp increase in pressure exerted by the shock wave. The overpressure at any given point is directly proportional to the density of the air in the wave.

- **Dynamic pressures**, i.e., drag exerted by the blast winds required to form the blast wave. These winds push, tumble and tear objects.

Most of the material damage caused by a nuclear air burst is caused by a combination of the high static overpressures and the blast winds. The long compression of the blast wave weakens structures, which are then torn apart by the blast winds. The compression, vacuum and drag phases together may last several seconds or longer, and exert forces many times greater than the strongest hurricane.

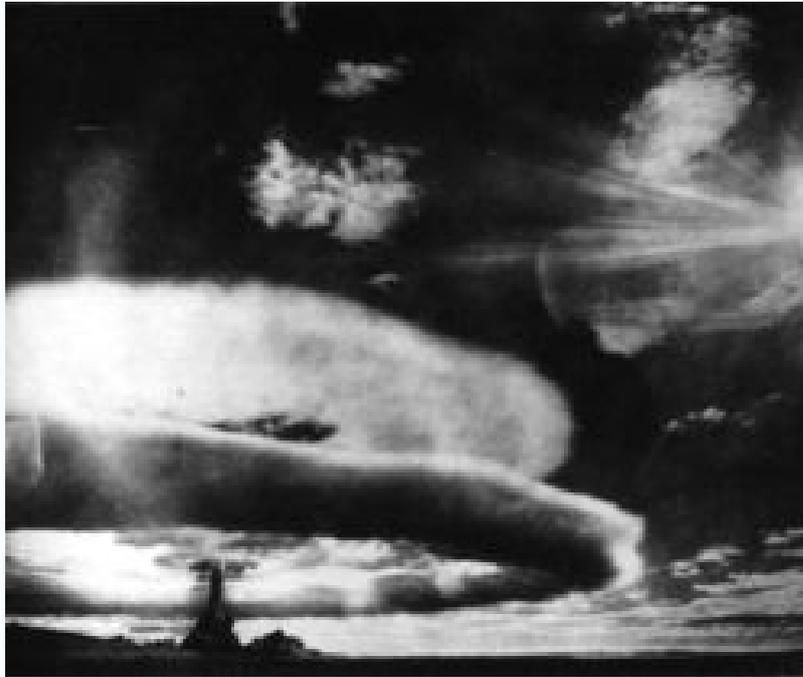

The mushroom cloud from the first "true" Soviet hydrogen bomb test in 1955.

Acting directly on the human body, the shock waves cause pressure waves through the tissues. These waves mostly damage junctions between tissues of different densities (bone and muscle) or the interface between tissue and air. Lungs and the abdominal cavity, which contain air, are particularly injured. The damage causes severe haemorrhaging or air embolisms, either of which can be rapidly fatal. The overpressure estimated to damage lungs is about 70 kPa. Some eardrums would probably rupture around 22 kPa (0.2 atm) and half would rupture between 90 and 130 kPa (0.9 to 1.2 atm).

**Blast Winds**: The drag energies of the blast winds are proportional to the cubes of their velocities multiplied by the durations. These winds may reach several hundred kilometers per hour.

**Thermal radiation**.

Nuclear weapons emit large amounts of electromagnetic radiation as visible, infrared, and ultraviolet light. The chief hazards are burns and eye injuries. On clear days, these injuries can occur well beyond blast ranges. The light is so powerful that it can start fires that spread rapidly in the debris left by a blast. The range of thermal effects increases markedly with weapon yield. Thermal radiation accounts for between 35-45% of the energy released in the explosion, depending on the yield of the device.

There are two types of eye injuries from the thermal radiation of a weapon:

Flash blindness is caused by the initial brilliant flash of light produced by the nuclear detonation. More light energy is received on the retina than can be tolerated, but less than is required for irreversible injury. The retina is particularity susceptible to visible and short wavelength infrared light, since this part of the electromagnetic spectrum is focused by the lens on the retina. The result is bleaching of the visual pigments and temporary blindness for up to 40 minutes.

When thermal radiation strikes an object, part will be reflected, part transmitted, and the rest absorbed. The fraction that is absorbed depends on the nature and color of the material. A thin material may transmit a lot. A light colored object may reflect much of the incident radiation and thus escape damage. The absorbed thermal radiation raises the temperature of the surface and results in scorching, charring, and burning of wood, paper, fabrics, etc. If the material is a poor thermal conductor, the heat is confined to the surface of the material.

Actual ignition of materials depends on how long the thermal pulse lasts and the thickness and moisture content of the target. Near ground zero where the energy flux exceeds 125 J/cm², what can burn, will. Farther away, only the most easily ignited materials will flame. Incendiary effects are compounded by secondary fires started by the blast wave effects such as from upset stoves and furnaces.

In Hiroshima, a tremendous fire storm developed within 20 minutes after detonation and destroyed many more buildings and homes. A fire storm has gale force winds blowing in towards the center of the fire from all points of the compass. It is not, however, a phenomenon peculiar to nuclear explosions, having been observed frequently in large forest fires and following incendiary raids during World War II.

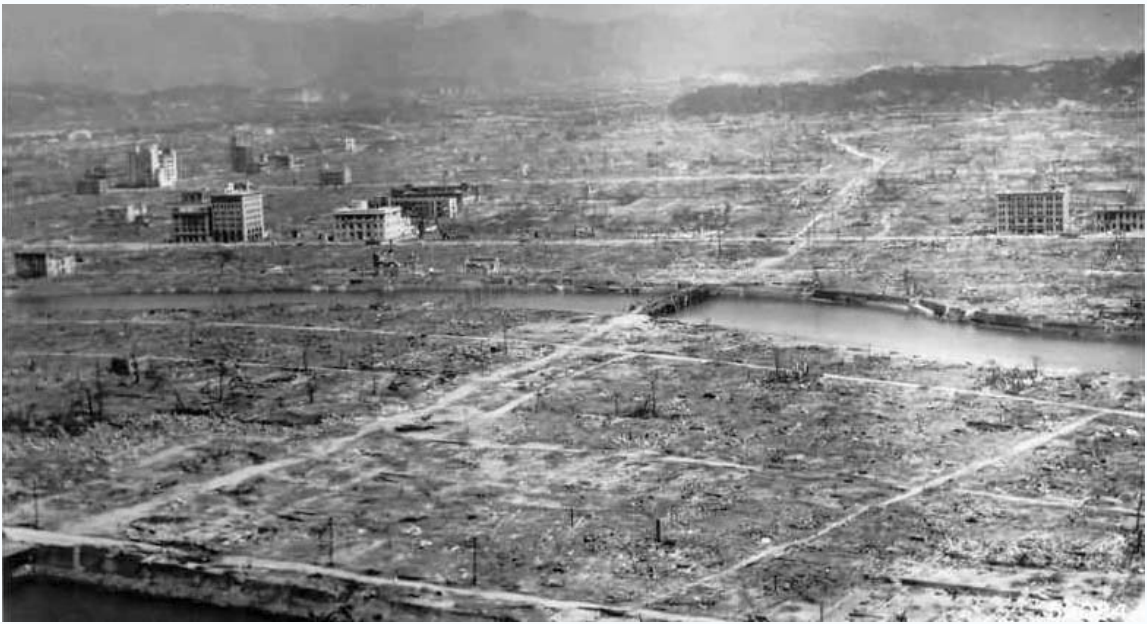

The desolation after the atomic bombing of Hiroshima in 1945

Because thermal radiation travels more or less in a straight line from the fireball (unless scattered) any opaque object will produce a protective shadow. If fog or haze scatters the light, it will heat things from all directions and shielding will be less effective, but fog or haze would also diminish the range of these effects.

**Indirect effect**. *Ionizing radiation*

About 5% of the energy released in a nuclear air burst is in the form of ionizing radiation: neutrons, gamma rays, alpha particles, and electrons moving at incredible speeds, but with different speeds that can be still far away from the speed of light (beta particles). The neutrons result almost exclusively from the fission and fusion reactions, while the initial gamma radiation includes that arising from these reactions as well as that resulting from the decay of short-lived fission products.

The intensity of initial nuclear radiation decreases rapidly with distance from the point of burst because the radiation spreads over a larger area as it travels away from the explosion. It is also reduced by atmospheric absorption and scattering.

The character of the radiation received at a given location also varies with distance from the explosion. Near the point of the explosion, the neutron intensity is greater than the gamma intensity, but with increasing distance the neutron-gamma ratio decreases. Ultimately, the neutron component of initial radiation becomes negligible in comparison with the gamma component. The range for significant levels of initial radiation does not increase markedly with weapon yield and, as a result, the initial radiation becomes less of a hazard with increasing yield. With larger weapons, above fifty kt (200 TJ), blast and thermal effects are so much greater in importance that prompt radiation effects can be ignored.

The neutron radiation serves to transmute the surrounding matter, often rendering it radioactive. When added to the dust of radioactive material released by the bomb itself, a large amount of radioactive material is released into the environment. This form of radioactive contamination is known as nuclear fallout and poses the primary risk of exposure to ionizing radiation for a large nuclear weapon.

*Electromagnetic pulse.*

Gamma rays from a nuclear explosion produce high energy electrons through Compton scattering. These electrons are captured in the earth's magnetic field, at altitudes between twenty and forty kilometers, where they resonate. The oscillating electric current produces a coherent electromagnetic pulse (EMP) which lasts about one millisecond. Secondary effects may last for more than a second.

The pulse is powerful enough to cause long metal objects (such as cables) to act as antennae and generate high voltages when the pulse passes. These voltages, and the associated high currents, can destroy unshielded electronics and even many wires. There are no known biological effects of EMP. The ionized air also disrupts radio traffic that would normally bounce off the ionosphere.

One can shield electronics by wrapping them completely in conductive mesh, or any other form of Faraday cage. Of course radios cannot operate when shielded, because broadcast radio waves can't reach them.

## Summary of the effects

The following table summarizes the most important effects of nuclear explosions under certain conditions.

| Effects | Explosive yield / Height of Burst | | | |
|---|---|---|---|---|
| | 1 kT / 200 m | 20 kT / 540 m | 1 MT / 2.0 km | 20 MT / 5.4 km |
| Blast—effective ground range *GR* / km | | | | |
| Urban areas almost completely leveled (20 PSI) | 0.2 | 0.6 | 2.4 | 6.4 |
| Destruction of most civil buildings (5 PSI) | 0.6 | 1.7 | 6.2 | 17 |
| Moderate damage to civil buildings (1 PSI) | 1.7 | 4.7 | 17 | 47 |
| **Thermal radiation—effective ground range *GR* / km** | | | | |
| Conflagration | 0.5 | 2.0 | 10 | 30 |
| Third degree burns | 0.6 | 2.5 | 12 | 38 |
| Second degree burns | 0.8 | 3.2 | 15 | 44 |
| First degree burns | 1.1 | 4.2 | 19 | 53 |
| **Effects of instant nuclear radiation—effective slant range[1] *SR* / km** | | | | |
| Lethal[2] total dose (neutrons and gamma rays) | 0.8 | 1.4 | 2.3 | 4.7 |
| Total dose for acute radiation syndrome[2] | 1.2 | 1.8 | 2.9 | 5.4 |

[1]) For the direct radiation effects the slant range instead of the ground range is shown here, because some effects are not given even at ground zero for some burst heights. If the effect occurs at ground zero the ground range can simply be derived from slant range and burst altitude.

[2]) "Acute radiation syndrome" corresponds here to a total dose of one gray, "lethal" to ten grays. Note that this is only a rough estimate since biological conditions are neglected here.

*Survivability*. This is highly dependent on factors such as proximity to the blast and the direction of the wind carrying fallout.

## 2. Possible types of nuclear wars

The possibility of using nuclear weapons in war is usually divided into two subgroups, each with different effects and potentially fought with different types of nuclear armaments.

The first, a ***limited nuclear war*** (sometimes *attack* or *exchange*), refers to a small scale use of nuclear weapons by one or more parties. A "limited nuclear war" would most likely consist of a limited exchange between two nuclear superpowers targeting each other's military facilities, either as an attempt to pre-emptively cripple the enemy's ability to attack as a defensive measure or as a prelude to an invasion by conventional forces as an offensive measure. It will also refer to a nuclear war between minor nuclear powers, who lack the ability to deliver a decisive strike. This term would apply to any limited use of nuclear weapons, which may involve either military or civilian targets.

The second, a ***full-scale nuclear war***, consists of large numbers of weapons used in an attack aimed at an entire country, including both military and civilian targets. Such an attack would seek to destroy the entire economic, social, and military infrastructure of a nation by means of an overwhelming nuclear attack. Such a nuclear war would be unlikely to remain contained between only the two countries involved, especially if either of the nuclear superpowers were involved.

Some Cold War strategists argued that a limited nuclear war could be possible between two heavily armed superpowers (such as the United States and the Soviet Union) and if so several predicted that a limited war could "escalate" into an all-out war. Others have called limited nuclear war "global nuclear holocaust in slow motion" arguing that once such a war took place others would be sure to follow over a period of decades, effectively rendering the planet uninhabitable in the same way that a "full-scale nuclear war" between superpowers would, only taking a much longer and more agonizing path to achieve the same result.

Even the happiest thoughts about the effects of a major nuclear exchange predict the death of millions of civilians within a very short amount of time; more pessimistic predictions argue that a full-scale nuclear war could bring about the extinction of the human race or its near extinction with a handful of survivors (mainly in remote areas) reduced to a pre-medieval quality of life and life expectancy for centuries after and cause permanent damage to most complex life on the planet, Earth's ecosystems, and the global climate. It is in this latter mode that nuclear warfare is usually alluded to as a doomsday scenario.

A third category, not usually included with the above two, is ***accidental nuclear war***, in which a nuclear war is triggered unintentionally. Possible scenarios for this have included malfunctioning early warning devices and targeting computers, deliberate malfeasance by rogue military commanders, accidental straying of planes into enemy airspace, reactions to unannounced missile tests during tense diplomatic periods, reactions to military exercises, mistranslated or miscommunicated messages, and so forth. A number of these scenarios did actually occur during the Cold War, though none resulted in a nuclear exchange. Many such scenarios have been depicted in popular culture, such as in the 1962 novel *Fail-Safe* and the 1964 film *Dr. Strangelove or: How I Learned to Stop Worrying and Love the Bomb*.

**Post-Cold War.** Although the collapse of the Soviet Union ended the Cold War and greatly reduced tensions between the United States and Russia, both nations remained in a "nuclear stand-off" due to the continuing presence of a significant number of warheads in both nations. Additionally, the end of the Cold War led the United States to become increasingly concerned with the development of nuclear technology by other nations outside of the former Soviet Union. In 1995, a branch of the U.S. Strategic Command produced an outline of forward-thinking strategies in the document "Essentials of Post-Cold War Deterrence".

The United Nations Organization's Disarmament Committee has announced that there are probably more than 16,000 strategic and tactical nuclear weapons ready for deployment and another 14,000 in storage. The U.S.A. has nearly 7,000 ready for action and 3,000 in storage and Russia (RF) has about 8,500 on hand and 11,000 in storage. China has, possibly, 400 nuclear weapons, Britain 400, France 350, India 95, and Pakistan 50. North Korea is confirmed as having nuclear weapons, though it is not known how many (a common estimate is between 1 and 10). Also, despite denials, Israel is also widely believed to have nuclear weapons, possibily as many as 200. NATO has stationed 480 U.S.A-built nuclear weapons in Belgium, the Netherlands, Italy, Germany, and Turkey, with several other countries in pursuit of an arsenal of their own.

A key development in nuclear warfare in the 21$^{st}$ Century has been the proliferation of nuclear weapons to the developing world, with Pakistan and India both publicly testing nuclear devices and North Korea conducting an underground nuclear test on October 9, 2006. The U.S. Geological Survey measured a 4.2 magnitude earthquake in the area where some form of testing occurred. It may have been a dud. Iran, meanwhile, has embarked on a nuclear program which, while officially for civilian purposes, has come under scrutiny by the United Nations and individual states.

Israel has been involved in wars with its neighbours on numerous occasions, and its small geographic size would mean that in the event of future wars the Israeli military might have very little time to react to a future invasion or other major threat; the situation could escalate to nuclear warfare very quickly in some scenarios. In addition, the fact that Iran appears to many observers to be in the process of developing a nuclear weapon has heightened fears of a nuclear conflict in the Middle East, either with Israel or with Iran's Sunni neighbours.

**Potential consequences of a regional nuclear war.** A study presented at the annual meeting of the American Geophysical Union in December 2006 asserted that even a small-scale, regional nuclear war could produce as many direct fatalities as all of World War II and disrupt the global climate for a decade or more. In a regional nuclear conflict scenario where two opposing nations in the subtropics would each use 50 Hiroshima-sized nuclear weapons (ca. 15 kiloton each) on major populated centers, the researchers estimated fatalities from 2.6 million to 16.7 million per country. Also, as much as five million tons of soot would be released, which would produce a cooling of several degrees Centigrade over large areas of North America and Eurasia, including most of the grain-growing regions. The cooling would last for years and could be "catastrophic" according to the researchers. This event-process is, sometimes, dubbed "Nuclear Winter".

**Sub-strategic use.** The above examples envisage nuclear warfare at a strategic level, i.e. total war. However, many nuclear powers are believed to have the ability to launch more limited engagements. The United Kingdom has reserved the possibility of launching a sub-strategic nuclear strike against an enemy, described by its Parliamentary Defence Select Committee as "the launch of one or a limited number of missiles against an adversary as a means of conveying a political message, warning or demonstration of resolve". This would see the deployment of strategic nuclear weapons in a very limited role rather than the battlefield exchanges of tactical nuclear weapons.

British Trident SSBN submarines are believed to carry some missiles for this purpose, potentially allowing a strike as low as one kiloton against a single target. Former Defence Secretary Malcolm Rifkind argued that this capacity offset the reduced credibility of fullscale strategic nuclear attack following the end of the Cold War *circa* 1990.

Commodore Tim Hare, former Director of Nuclear Policy at the UK's Ministry of Defence, has described it as offering the Government "an extra option in the escalatory process before it goes for an all-out strategic strike which would deliver unacceptable damage".

However, this sub-strategic capacity has been criticised as potentially increasing the acceptability of using nuclear weapons. The related consideration of new generations of limited yield battlefield nuclear weapons by the United States has also alarmed anti-nuclear groups, who believe it will make the use of nuclear weapons more acceptable.

**Nuclear terrorism.** Early in the 21st century, concerns began that "rogue states" such as North Korea and Iran could acquire or manufacture nuclear weapons and use them to attack a foe indirectly through terrorism. Nuclear terrorism by non-state organizations may be more likely, as states possessing nuclear weapons are susceptible to retaliation in kind. Geographically dispersed and mobile terrorist organizations are not so easy to discourage by the threat of retaliation. Furthermore, while the collapse of the Soviet Union ended the Cold War, it greatly increased the risk that former Soviet nuclear weapons might become available on the black market. Indeed, it has been alleged that several suitcase-size nuclear fission bombs might have been available. Using such a weapon as a foundation, a terrorist might even create a salted bomb capable of dispersing radioactive contamination over a large area, killing a greater number of people than the explosion itself. Also, "Dirty Bombs" can be made with common explosives used to spread radioactive particles, poisoning and sickening people directly affected by the fallout rather than the actual bomb shockwave.

## 3. Current methods of protection from the nuclear warheads and their disadvantages

The **Strategic Defense Initiative** (SDI) was a proposal by U.S. President Ronald Reagan on March 23, 1983 to use ground-based and space-based systems to protect the United States from attack by strategic **nuclear ballistic missiles**. The initiative focused on strategic defense rather than the prior strategic offense doctrine of **mutual assured destruction** (MAD).

Though it was never fully developed or deployed, the research and technologies of SDI paved the way for some **anti-ballistic missile** systems of today. The Strategic Defense Initiative Organization (SDIO) was set up in **1984** within the **United States Department of Defense** to oversee the Strategic Defense Initiative. (Amongst its negative critics, it gained the popular name *Star Wars* after the 1977 Hollywood-made movie by **George Lucas**.) Under the administration of President **Bill Clinton** in **1993**, its name was changed to the **Ballistic Missile Defense Organization** (BMDO) and its emphasis was shifted from national missile defense to theater missile defense; from global to regional coverage. BMDO was later renamed to the **Missile Defense Agency**. This article covers defense efforts under the SDIO.

SDI was not the first U.S.A. defensive system against nuclear ballistic missiles. In the 1960s, **The Sentinel Program** was designed and developed to provide a limited defensive capability, but was never deployed. Sentinel technology was later used in the **Safeguard Program**, briefly deployed to defend one U.S. location. In the 1970s the Soviet Union deployed a missile defense system, still operational today, which defends Moscow and nearby missile sites.

SDI is unique from the earlier U.S.A. and Soviet missile defense efforts. It envisioned using space-oriented basing of defensive systems vs solely ground-launched interceptors. It also initially had the ambitious goal of a near total defense against a massive sophisticated **ICBM** attack, vs previous systems which were limited in defensive capacity and geographic coverage.

In 1984, the Strategic Defense Initiative Organization (SDIO) was established to oversee the program, which was headed by Lt. General **James Alan Abrahamson**, USAF, a past Director of the NASA **Space Shuttle program**. Research and development initiated by the SDIO created significant technological advances in computer systems, component miniaturization, sensors and missile systems that form the basis for current systems.

Initially, the program focused on large scale systems designed to defeat a Soviet offensive strike. However, as the threat diminished, the program shifted towards smaller systems designed to defeat limited or accidental launches.

By 1987, the SDIO developed a national missile defense concept called the Strategic Defense System Phase I Architecture. This concept consisted of ground and space based sensors and weapons, as well as a central battle management system. The **ground-based systems operational today** trace their roots back to this concept.

In his 1991 **State of the Union Address George H. W. Bush** shifted the focus of SDI from defense of North America against large scale strikes to a system focusing on theater missile defense called Global Protection Against Limited Strikes (GPALS).

In 1993, the **Clinton** administration, further shifted the focus to ground-based interceptor missiles and theater scale systems, forming the **Ballistic Missile Defense Organization** (BMDO) and closing the SDIO. Ballistic missile defense has been revived by the **George W. Bush** administration as the **National Missile Defense** and Ground-based Midcourse Defense.

SDI included the following Programs:

- Ground-based programs
    - 1 Extended Range Interceptor (ERINT)
    - 2 Homing Overlay Experiment (HOE)

- - o   3 Exo-atmospheric Reentry-vehicle Interception System (ERIS)
- Directed-energy weapon (DEW) programs
  - o   1 X-ray laser
  - o   2 Chemical laser
  - o   3 Neutral Particle Beam
  - o   4 Laser and mirror experiments
  - o   5 Hypervelocity Rail Gun (CHECMATE)
- Space-based programs
  - o   1 Space-Based Interceptor (SBI)
  - o   2 Brilliant Pebbles
- Sensor programs
  - o   1 Boost Surveillance and Tracking System (BSTS)
  - o   2 Space Surveillance and Tracking System (SSTS)
  - o   3 Brilliant Eyes
  - o   4 Other sensor experiments

. There was a lot of criticism of non efficiency SDI.Another criticism of SDI was that it would not be effective against non-space faring weapons, namely cruise missiles, bombers, and non-conventional delivery methods such as delivery via commercial naval vessels. This latter method in particular would be attractive to terrorists and rogue states as it would be inexpensive, difficult to trace, and technologically undemanding.

   The USA Government spent hundreds billions of dollars but the USA does not have SDI now after 24 years R&D.

## Anti-ballistic missile.

An **anti-ballistic missile** (ABM) is a missile designed to counter ballistic missiles. A ballistic missile is used to deliver nuclear, chemical, biological or conventional warheads in a ballistic flight trajectory. The term "anti-ballistic missile" describes any antimissile system designed to counter ballistic missiles. However the term is more commonly used for ABM systems designed to counter long range, nuclear-armed Intercontinental ballistic missiles (ICBMs).

Only two ABM systems have previously been operational against ICBMs, the U.S. Safeguard system, which utilized the LIM-49A Spartan and Sprint missiles, and the Russian A-35 anti-ballistic missile system which used the Galosh interceptor, each with a nuclear warhead themselves. Safeguard was only briefly operational; the Russian system has been improved and is still active, now called A-135 and using two missile types, Gorgon and Gazelle, both with conventional warheads. However the U.S. Ground-Based Midcourse Defense (GMD, previously called NMD) system has recently reached initial operational capability. It does not have an explosive charge, but launches a kinetic projectile.

Three shorter range tactical ABM systems are currently operational: the U.S. Army Patriot, U.S. Navy Aegis combat system/Standard SM-3, and the Israeli Arrow. The longer-range U.S. Terminal High Altitude Area Defense (THAAD) system is scheduled for deployment in 2011. In general short-range tactical ABMs cannot intercept ICBMs, even if within range. The tactical ABM radar and performance characteristics do not allow it, as an incoming ICBM warhead moves much faster than a tactical missile warhead. However it is possible the higher performance THAAD missile could be upgraded to intercept ICBMs.

Latest versions of the U.S. Hawk missile have a limited capability against tactical ballistic missiles, but is usually not described as an ABM. Similar claims have been made about the Russia's long-range surface-to-air S-300 and S-400 series.

**Israel ABM**. The Arrow research effort was first co-funded by the U.S.A. and Israel on May 6, 1986.

The Arrow ABM system was designed and constructed in Israel with financial support by the United States in a multi-billion dollar development program called "Minhelet Homa" with the participation of companies such as the Israel Military Industries, Tadiran and Israel Aerospace Industries.

In 1998 the Israeli military conducted a successful test of their Arrow ABM. Designed to intercept incoming missiles travelling at up to 2 mile/s (3 km/s), the Arrow is expected to perform much better than the Patriot did in the Gulf War. On July 29, 2004 Israel and the United States carried out joint experiment in the USA, in which the Arrow was launched against a real Scud missile. The experiment was a success, as the Arrow destroyed the Scud with a direct hit. In December 2005 the system was successfully deployed in a test against a replicated Shahab-3 missile. This feat was repeated on February 11, 2007.

**Anti-satellite weapons** (ASATs)

These are space weapons designed to destroy satellites for strategic military purposes. Currently, only the USA, the former USSR and the People's Republic of China are known to have developed these weapons, with India claiming the technical capability to develop such weapons. On January 11, 2007, China destroyed an old orbiting weather satellite, the world's first test since the 1980s.

**ASAT in the era of Strategic Defense.** The era of the Strategic Defense Initiative (proposed in 1983) focussed primarily on the development of systems to defend against nuclear warheads, however, some of the technologies developed may be useful also for antisatellite use.

After the Soviet Union collapsed, there were proposals to use this aircraft as a launch platform for lofting commercial and science packages into orbit. Recent political developments (see below) may have seen the reactivation of the Russian Air-Launched ASAT program, although there is no confirmation of this as yet.

The Strategic Defense Initiative gave the U.S.A. and Russian ASAT programs a major boost; ASAT projects were adapted for ABM use and the reverse was also true. The initial US plan was to use the already developed MHV as the basis for a space based constellation of about 40 platforms deploying up to 1,500 kinetic interceptors. By 1988 the US project had evolved into an extended four stage development. The initial stage would consist of the Brilliant Pebbles defense system, a satellite constellation of 4,600 kinetic interceptors (KE ASAT), of 100 lb (45 kg) each, in Low Earth orbit, and their associated tracking system. The next stage would deploy the larger platforms and the following phases would include the laser and charged particle beam weapons that would be developed by that time from existing projects such as MIRACL. The first stage was intended to be completed by 2000 at a cost of around $125 billion.

Research in the US and Russia was proving that the requirements, at least for orbital based energy weapon systems, were, with available technology, close to impossible. Nonetheless, the strategic implications of a possible unforeseen breakthrough in technology forced the USSR to initiate massive spending on research in the 12th Five Year Plan, drawing all the various parts of the project together under the control of GUKOS and matching the US proposed deployment date of 2000.

Both countries began to reduce expenditure from 1989 and the Russian Federation unilaterally discontinued all SDI research in 1992. Research and Development (both of ASAT systems and other space based/deployed weapons) has, however reported to have be been resumed under the government of Vladimir Putin as a counter to renewed US Strategic Defense efforts post Anti-Ballistic Missile Treaty. However the status of these efforts, or indeed how they are being funded through National Reconnaissance Office projects of record, remains unclear. The U.S. has begun working on a number of programs which could be foundational for a space-based ASAT. These programs include the Experimental Spacecraft System (XSS 11), the Near-Field Infrared Experiment (NFIRE), and the space-based interceptor (SBI).

## Recent developments

On 14 October 2002, a ground based interceptor launched from the Ronald Reagan Ballistic Missile Defense Site destroyed a mock warhead 225 km above the Pacific. The test included three decoy balloons.

On 16 December 2002 President George W. Bush signed National Security Presidential Directive 23 which outlined a plan to begin deployment of operational ballistic missile defense systems by 2004. The following day the U.S.A. formally requested from the UK and Denmark use of facilities in Fylingdales, England, and Thule, Greenland, respectively, as a part of the NMD program. The projected cost of the program for the years 2004 to 2009 will be $53 billion, making it the largest single expenditure in America's defence department budget.

Since 2002, the US has been in talks with Poland and other European countries over the possibility of setting up a European base to intercept long-range missiles. A site similar to the US base in Alaska would help protect the US and Europe from missiles fired from the Middle East or North Africa. Poland's prime minister Kazimierz Marcinkiewicz said in November 2005 he wanted to open up the public debate on whether Poland should host such a base.

In 2002, NMD was changed to Ground-Based Midcourse Defense (GMD), to differentiate it from other missile defense programs, such as space-based, sea-based, and defense targeting the boost phase and the reentry phase (see flight phases).

On 22 July 2004, the first ground-based interceptor was deployed at Ft. Greely, Alaska (63.954° N 145.735° W). By the end of 2004, a total of six had been deployed at Ft. Greely and another two at Vandenberg Air Force Base, California. Two additional were installed at Ft. Greely in 2005. The system will provide "rudimentary" protection.

On 15 December 2004, an interceptor test in the Marshall Islands failed when the launch was aborted due to an "unknown anomaly" in the interceptor, 16 minutes after launch of the target from Kodiak Island, Alaska.

"I don't think that the goal was ever that we would declare it was operational. I think the goal was that there would be an operational capability by the end of 2004," Larry DiRita said on 13 January 2005 at a Pentagon press conference. However, the problem is and was funding "There has been some expectation that there will be some point at which it is operational and not something else these expectations are not unknown, if Congress pours more attention and funding to this system, it can be operational relatively quick."

On 18 January 2005, the Commander, United States Strategic Command issued direction to establish the Joint Functional Component Command for Integrated Missile Defense. JFCC IMD, once activated, will develop desired characteristics and capabilities for global missile defense operations and support for missile defense.

On 14 February 2005, another interceptor test failed due to a malfunction with the ground support equipment at the test range on Kwajalein Island, not with the interceptor missile itself.[6]

On 24 February 2005, the Missile Defense Agency, testing the Aegis Ballistic Missile Defense System, successfully intercepted a mock enemy missile. This was the first test of an operationally configured Standard missile 3 interceptor and the fifth successful test intercept using this system. On 10 November 2005, the USS *Lake Erie* detected, tracked, and destroyed a mock two-stage ballistic missile within two minutes of the ballistic missile launch.

On 1 September 2006, the Ground-Based Midcourse Defense System was successfully tested. An interceptor was launched from Vandenberg Air Force Base to hit a target missile launched from Alaska, with ground support provided by a crew at Colorado Springs. This test was described by Missile Defense Agency director Lieutenant General Trey Obering as "about as close as we can come to an end-to-end test of our long-range missile defense system."

Deployment of the Sea-based X-band Radar system is presently underway.

On 24 February 2007, The Economist, a magazine published in the UK, reported that the United States ambassador to NATO, Victoria Nuland, had written to her fellow envoys to advise them regarding the various options for missile-defence sites in Europe. She also confirmed that "The United States has also been discussing with the UK further potential contributions to the system."

In February 2007 US started formal negiotiations with Poland and Czech Republic concerning construction of missile shield installations in those countries for a Ground-Based Midcourse Defense System. According to press reports Czech republic agreed to host a missile defence radar on its territory while a base of missile interceptors is supposed to be built in Poland. The objective is reportedly to protect most of Europe from long-range missile strikes from Iran.

The Ustka-Wicko base of Polish Army is mentioned as a possible site of US missile interceptors.

## Criticism. Report of the American Physical Society

There has been controversy among experts about whether it is technically feasible to build an effective missile defense system. One technical criticism came from U.S.A. physicists and culminated in the publication of a critical study on the subject by the American Physical Society (APS).

This study focused on the feasibility of intercepting missiles in the boost phase, which the current NMD system does not attempt. The study found it might be possible to develop a limited system capable of destroying a liquid-fuel propelled ICBM during the boost phase. This system could also possibly destroy some solid-propellant missiles from Iran, but not those from North Korea, because of differences in the boost time and range to target. However, there is a trend toward using solid-fueled ICBMs which are harder to intercept during boost phase.

Using orbital launchers to provide a reliable boost-phase defense against solid fuel missiles from Iran or North Korea was found to require at least 1,600 interceptors in orbit. Intercepting liquid-fueled missiles would require 700 interceptors. Using two or more interceptors per target would require many more orbital launchers.

The only boost phase system the U.S.A. contemplates for near term use is the Airborne laser (ABL). The study found the ABL possibly capable of intercepting missiles if within 300 km for solid fuel missiles or 600 km for liquid fuel missiles, however solid fuel missiles are more resistant to damage.

While the APS report did not address the current U.S. mid-course NMD system, it concluded that were the U.S.A. in the future to develop a boost-phase ABM defense, there could be significant technical problems limiting effectiveness.

## Summary.

The cuurent defense systems are very complex, request very high technology and VERY EXPENSIVE. Only high industriel and very rich contries can develop or buy these systems. In many cases they cost in a lot of times more then the attack systems (strategic missiles and nuclear warheads), expecially the defence system from small rockets (as Qassam) or Mortar shells.

*Acknowledgement*: Some data in this work is garnered from Wikipedia under the Creative Commons License.

## Description of AB-Dome and Innovations

The author here offers a new kind of city-wide protection system against incoming warheads carrying a nuclear weapon. The AB-Dome is described in the author's works [1]-[4].

His idea is a thin dome covering a city with that is a very transparent film 2 (Fig.1). The film has thickness 0.05 – 0.3 mm. One is located at high altitude (5 - 20 km). The film is supported at this altitude by a small additional air pressure produced by ground ventilators. That is connected to Earth's ground by managed cables 3. The film may have a controlled transparency option. The system can have the second lower film 6 with controlled reflectivity, a further option.

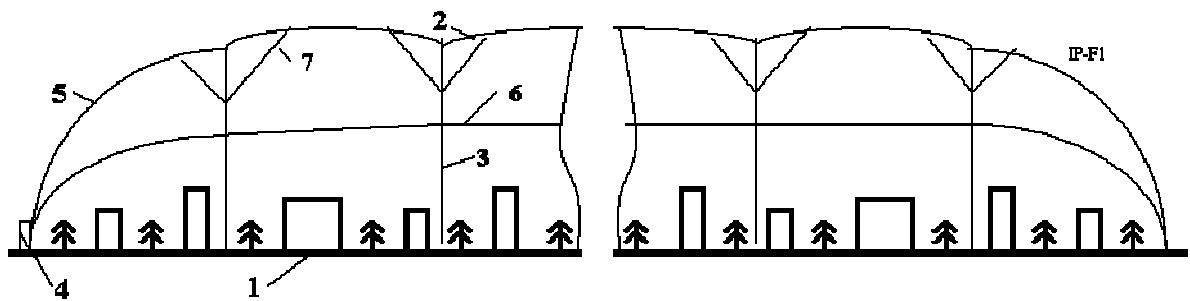

**Fig. 1.** Film AB-Dome for big city. *Notations*: 1 - area, 2 - thin film cover with a control clarity (option), 3 – control support cable height is 10 – 15 km), 4 - exits and ventilators, 5 - border section, 6 – the second (lower) control reflectivity film cover (option), 7 – additional support cables.

The small additional pressure creates a signufically lift force. For example, the additional pressure only $p$ = 0.01 atm produces the lift force about 100 kg/m$^2$. At altitute $H$ = 7 km this force is more 40 kg/m$^2$; at altitude $H$ =10 km, the force is 26 kg/m$^2$; at altitude $H$ =15 km, the force is 12 kg/m$^2$. The support cable has a weigth about 5-10 kg/m$^2$, the 1 m$^2$ of film weights less 0.05 - 0.5 kg (for example, Kevlar film of thickness 0.2 mm has the weight 0.3 kg/m$^2$). That means every square meter of dome can support a useful load 10 - 30 kg. At high altitude the useful load decreses, but if it is needed, it can be increased by increasing the interior air pressure.

The top film may have the cheap pebbles (stones) 8 (Fig.2) on the upper surface. If the pebble has mass 0.5 kg and a step 0.5 m, the total stone weight will be 2 kg/m$^2$.

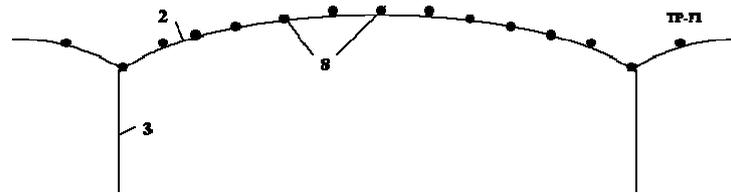

**Fig. 2**. The top film is armored by strong stones (pebbles) 8.

**Wartime**. The offered protection defends in the following way. The smallest space warhead has a minimum cross-section area 1 m² and a huge speed 3 – 5 km/s. The warhead gets a blow and overload from film (mass about 0.5 kg). This overload is 500 – 1500g and destroys the warhead (see computation below). Warhead also gets an overpowering blow from 2 -5 (every mass is 0.5 - 1 kg) of the strong stones. Relative (about warhead) kinetic energy of every stone is about 8 millions of Joules! (It is in 2-3 more than energy of 1 kg explosive!). The film destroys the high speed warhead (aircraft, bomber, wing missile) especially if the film will be armored by stone as it is described in Fig. 2, above.

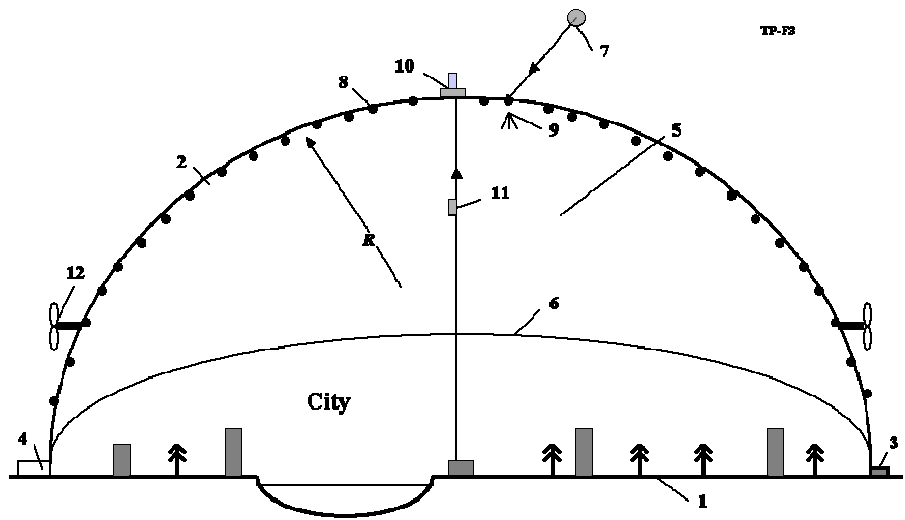

**Fig.3.** Spherical AB-Dome for big city against thermonuclear warheads, strategic rockets, missiles, aviation, chemical and biological weapons. *Notations*: 1 – protected area; 2 – thin film; 3 – ventilator (air pump); 4 – exit; 5 –spherical thin film AB-Dome; 6 – control reflectivity thin film (optional); 7 – nuclear warhead, rocket, missile, bomb or aviation; 8 – strong stones; 9 – fragments of destroyed warhead, rocket, missile or aircraft; 10 – TV, communication, telescope, locator, tourists; 11 – elevator; 12 – windmills.

*Note*: a nuclear bomb is very different from a conventional explosive bomb. Small inaccuracy in the connection of nuclear fuel parts (in the location or times) makes the bomb a pieces of junk or decreases it efficiency by hundreds times! In most cases, the destroyed nuclear warhead (bomb) will fall to Earth. The nuclear fuel can be taken from it (as part of any widespread decontamination effort) and then, possibly, used for making the defender's own nuclear bomb since the city and nation affected will still have operating weapons construction infrastructure! The basic design for a nuclear bomb is well known, and the main problem is to get the nuclear fuel (explosive). That means the defending country receives the nuclear weapon for the deserved punishment or attack enemy.

The optimal (for damage) height of nuclear explosion is 50-500 m (it depends entirely on the nuclear bomb's explosive yield. The more powerful the bomb, the height ought to be accordingly higher). Low height allows getting the strong air blast and neutrons reach a ground and produce a lot of radioactive isotopes (dust). If the nuclear bomb explodes at altitude 7 -15 km the bomb efficiency decreases by hundreds and thousands times (see theoretical, and computation sections below). Why? The nuclear bomb has two main effects: air blast and thermal radiation. Both effects decrease in third order from blast site distance. That means if distance increases in 10 times, the efficiency decreases in $10^3$ =1000 times! For example, the bomb has efficiency 0.5 km. That is exploded at altitude 10 km. Their efficiency decreases in $20^3$ = 8000 times! Add to this that a relative air density at altitude 10 km is 4 times less than the air density at sea level, the air blast then decreases additionally by 4 times (total decrease will be 32,000 times!).

The same situation occurs with post-detonation thermal radiation. In above example, it decreases by 8000 times. The lower film layer will automatically increase their reflectivity and one can decrease the thermal radiation in additionally about 10 times (total decreasing will be in 80,000 times!). Internal AB-Dome white-colored water vapor clouds (located at altitude 2-4 km) additionally decreases the thermal radiation present.

The lower film automatically increases the light reflectivity and additionally decreases the thermal radiation in 10 - 20 times. The bomb exploded at high altitude (in rare atmosphere) has significantly less air blast and small radioisotope contamination (the neutrons don't reach the ground).

The city saves the resident life and gets a minimal damage.

In **peacetime,** the offered AB-Dome produces a fine warm climate (weather) in a covered city. The films can have a controlled transparency and reflectivity. That allows provision of different solar heating conditions within the shielded city. These gigantic protective covers are composed of a cheap film having liquid crystal and conducting layers. The clarity of them is controlled by electric voltage. They can pass or block the incidental sunlight (or parts of the solar spectrum) and pass or blockade the Earth radiation. The outer and inside radiations have different wavelengths. That makes it possible to control of them separately and to control heating into (and re-radiation from) the Earth's surface. In conventional conditions about 50% of the solar energy reaches the Earth surface. The most part is reflected back to outer space by the white clouds. In our closed system the clouds (and rain, or at least condensation based dripping) will occur at night when the temperature is low. That means the many cold regions (Alaska, Siberia) may absorb more net solar energy and became, within the bubbles, lands with a temperate or sub-tropic climate. That also means the Sahara desert can be a prosperous area with a fine climate and with closed-loop water cycle.

The building of a film dome is very easy. Don't think of a present science fiction dome city made of impervious thick crystal geodesic panels. We simply spread out the film over Earth surface, turn on the pumping fans, and the film is gradually and controllably raised by air over-pressure inside to needed altitudes, limited by the support cables. Damage to the film is not a major trouble because the additional air pressure is very small and propeller pumps compensate for any air leakage. Unlike in an outer space colony or thin-atmosphere planetary colony (for example, Mars), the outside air is friendly and at worst we lose some heat (or cold) and water vapor.

The other advantages of the suggested method include the possibility to paint pictures on the sky (dome), to show films on the artificial sky by projector, to suspend illuminations, decorations, and air tramways and any other utilities and conveniences (and engineering works) from this new type of roofing.

Long distance aircraft fly at altitude between 8 - 11 km and our dome (less 7- 10 km) does not trouble them unless the dome is built on the edge of a glide path (inbound) or outbound departure path to an airport! The restraining cables will have safety illumination lights (red, flashing, in a string) and internal helicopters will take normal precautions in avoiding contact with them.

More detail the offered AB-Dome is described in [1]-[4] and [13]. Additional information is repeated below.

Our design for the AB-Dome is presented in Figures 1-3, which include the thin inflated film dome. The **innovations** are listed here: (1) the construction is air-inflatable; (2) each dome is fabricated with very thin, transparent film (thickness is 0.05 to 0.3 mm, implying under 150-500 tons a square kilometer) having the control clarity quality without rigid supports; (3) the enclosing film can have (option) two conductivity layers plus a liquid crystal layer between them which changes its clarity, color and reflectivity under an electric voltage; (4) the bound section of dome has a hemisphere form. The air pressure is more in these sections and they protect the central sections from outer wind.

Figs. 1-3 illustrate the thin transparent control dome cover we envision. The inflated textile shell—technical "textiles" can be woven or non-woven (films)—embodies the innovations listed: (1) the film is very thin, approximately 0.1 to 0.3 mm. A film this thin has never before been used in a major building; (2) the film has two strong nets, with a mesh of about 0.1 × 0.1 m and $a = 1 \times 1$ m, the threads are about 0.5 mm for a small mesh and about 1 mm for a big mesh. The net prevents the watertight and airtight film covering from being damaged by vibration; (3) the film incorporates a tiny electrically conductive wire net with a mesh about 0.1 x 0.1 m and a line width of about 100 $\mu$ and a thickness near 10 $\mu$. The wire net is electric (voltage) control conductor. It can inform the dome supervisors concerning the place and size of film damage (tears, rips, etc.); (4) the film may be twin-layered with the gap — $c = 1$ m and $b = 2$ m—between covering's layers for heat saving. In polar regions this multi-layered low height covering is the main means for heat insulation and puncture of one of the layers won't cause a loss of shape because the film's second layer is unaffected by holing; (5) the airspace in the dome's covering can be partitioned, either hermetically or not; and (6) part of the covering can have a very thin shiny aluminum coating that is about 1$\mu$ for reflection of unnecessary solar radiation in equatorial or polar regions [1]-[4].

## Theory and computation (estimation) of AB-Dome

### a) General information

Our dome cover (film) has 2 layers (figs. 1,3): top transparent layer 2, located at a maximum altitude (up 5 -20 km), and lower transparant layer 4 having control reflectivity, located at altitude of 1-3 km (option). Upper transparant cover has thickness about 0.05 – 0.3 mm and supports the protection strong stones (rebbles) 8. The stones have a mass 0.2 – 1 kg and locate the step about 0.5 m.

**Brief information about cover film.** If we want to control temperature in city, the top film must have some layers: transparent dielectric layer, conducting layer (about 1 - 3 $\mu$), liquid crystal layer (about 10 - 100 $\mu$), conducting layer (for example, $SnO_2$), and transparant dielectric layer. Common thickness is 0.05 - 0.5 mm. Control voltage is 5 - 10 V. This film may be produced by industry relatively cheaply.

   **1. Liquid crystals** (LC) are substances that exhibit a phase of matter that has properties between those of a conventional liquid, and those of a solid crystal.

   Liquid crystals find wide use in liquid crystal displays (LCD), which rely on the optical properties of certain liquid crystalline molecules in the presence or absence of an electric field. The electric field can be used to make a pixel switch between clear or dark on command. Color LCD systems use the same technique, with color filters used to generate red, green, and blue pixels. Similar principles can be used to make other liquid crystal based optical devices. Liquid crystal in fluid form is used to detect electrically generated hot spots for failure analysis in the semiconductor industry. Liquid crystal memory units with extensive capacity were used in Space Shuttle navigation equipment. It is also worth noting that many common fluids are in fact liquid crystals. Soap, for instance, is a liquid crystal, and forms a variety of LC phases depending on its concentration in water.

   The conventional control clarity (transparancy) film reflected a superfluos energy back to space. If film has solar cells that converts the superfluos solar energy into electricity.

   **2. Transparency**. In optics, transparency is the material property of allowing light to pass through. Though transparency usually refers to visible light in common usage, it may correctly be used to refer to any type of radiation. Examples of transparent materials are air and some other gases, liquids such as water, most glasses, and plastics such as Perspex and Pyrex. Where the degree of transparency varies according to the wavelength of the light. From electrodynamics it results that only a vacuum is really transparent in the strict meaning, any matter has a certain absorption for electromagnetic waves. There are transparent glass walls that can be made opaque by the application of an electric charge, a technology known as electrochromics. Certain crystals are transparent because there are straight lines through the crystal structure. Light passes unobstructed along these lines. There is a complicated theory "predicting" (calculating) absorption and its spectral dependence of different materials.

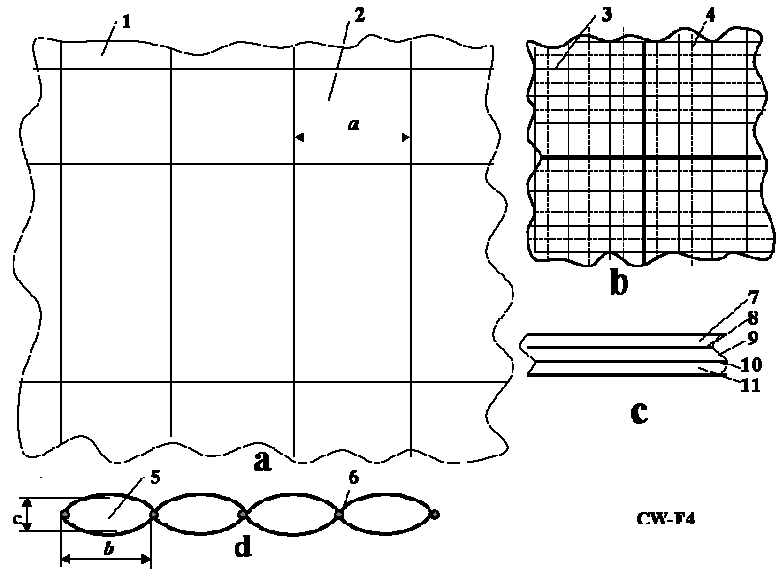

**Fig. 4.** Design of covering membrane. *Notations*: (a) Big fragment of cover with control clarity (reflectivity, carrying capacity) and heat conductivity; (b) Small fragment of cover; (c) Cross-section of cover (film) having 5 layers; (d) Longitudinal cross-section of low height cover for cold and hot regions (optional); 1 - cover; 2 - mesh; 3 - small mesh; 4 - thin electric net; 5 - cell of cover; 6 - tubes;: 7 - transparant dielectric layer, 8 - conducting layer (about 1 - 3 μ), 9 - liquid crystal layer (about 10 - 100 μ), 10 - conducting layer, and 11 - transparent dielectric layer. Common thickness is 0.1 - 0.5 mm. Control voltage is 5 - 10 V.

   **3. Electrochromism** is the phenomenon displayed by some chemical species of reversibly changing color when a burst of charge is applied.
  One good example of an electrochromic material is polyaniline which can be formed either by the electrochemical or chemical oxidation of aniline. If an electrode is immersed in hydrochloric acid which contains a small concentration of aniline, than a film of polyaniline can be grown on the electrode. Depending on the redox state, polyaniline can either be pale yellow or dark green/black. Other electrochromic materials that have found technological application include the viologens and polyoxotungstates. Other electrochromic materials include tungsten oxide ($WO_3$), which is the main chemical used in the production of electrochromic windows or smart windows.
  As the color change is persistent and energy need only be applied to effect a change, electrochromic materials are used to control the amount of light and heat allowed to pass through windows ("smart windows"), and has also been applied in the automobile industry to automatically tint rear-view mirrors in various lighting conditions. Viologen is used in conjunction with titanium dioxide ($TiO_2$) in the creation of small digital displays. It is hoped that these will replace LCDs as the viologen (which is typically dark blue) has a high contrast to the bright color of the titanium white, therefore providing a high visibility of the display.
  **4. Film and cable properties** [16]-[19]**.** Artificial fibers are currently being manufactured, which have tensile strengths of 3-5 times more than steel and densities 4-5 times less than steel. There are also experimental fibers (whiskers) which have tensile strengths 30-100 times more than a steel and densities 2 to 5 times less than steel. For example, in the book [16] p.158 (1989), there is a fiber (whisker) $C_D$, which has a tensile strength of $\sigma = 8000$ kg/mm$^2$ and density (specific gravity) of $\gamma = 3.5$ g/cm$^3$. If we use an estimated strength of 3500 kg/mm$^2$ ($\sigma = 7 \cdot 10^{10}$ N/m$^2$, $\gamma = 3500$ kg/m$^3$), than the ratio is $\gamma/\sigma = 0.1 \times 10^{-6}$ or $\sigma/\gamma = 10 \times 10^6$. Although the described (1989) graphite fibers are strong ($\sigma/\gamma = 10 \times 10^6$), they are at least still ten times weaker than theory predicts. A steel fiber has a tensile strength of 5000 MPA (500 kg/sq. mm), the theoretical limit is 22,000 MPA (2200

kg/mm$^2$)(1987); the polyethylene fiber has a tensile strength 20,000 MPA with a theoretical limit of 35,000 MPA (1987). The very high tensile strength is due to its nanotube structure [19].

Apart from unique electronic properties, the mechanical behavior of nanotubes also has provided interest because nanotubes are seen as the ultimate carbon fiber, which can be used as reinforcements in advanced composite technology. Early theoretical work and recent experiments on individual nanotubes (mostly MWNT's, Multi Wall Nano Tubes) have confirmed that nanotubes are one of the stiffest materials ever made. Whereas carbon-carbon covalent bonds are one of the strongest in nature, a structure based on a perfect arrangement of these bonds oriented along the axis of nanotubes would produce an exceedingly strong material. Traditional carbon fibers show high strength and stiffness, but fall far short of the theoretical, in-plane strength of graphite layers by an order of magnitude. Nanotubes come close to being the best fiber that can be made from graphite.

For example, whiskers of Carbon nanotube (CNT) material have a tensile strength of 200 Giga-Pascals and a Young's modulus over 1 Tera Pascals (1999). The theory predicts 1 Tera Pascals and a Young's modulus of 1-5 Tera Pascals. The hollow structure of nanotubes makes them very light (the specific density varies from 0.8 g/cc for SWNT's (Single Wall Nano Tubes) up to 1.8 g/cc for MWNT's, compared to 2.26 g/cc for graphite or 7.8 g/cc for steel). Tensile strength of MWNT's nanotubes may reach 150 GPa.

Specific strength (strength/density) is important in the design of the systems presented in this paper; nanotubes have values at least 2 orders of magnitude greater than steel. Traditional carbon fibers have a specific strength 40 times that of steel. Since nanotubes are made of graphitic carbon, they have good resistance to chemical attack and have high thermal stability. Oxidation studies have shown that the onset of oxidation shifts by about $100^0$ C or higher in nanotubes compared to high modulus graphite fibers. In a vacuum, or reducing atmosphere, nanotube structures will be stable to any practical service temperature (in vacuum up $2800\,^{o}$C. in air up $750^{o}$C).

In theory, metallic nanotubes can have an electric current density (along axis) more than 1,000 times greater than metals such as silver and copper. Nanotubes have excellent heat conductivity along axis up 6000 W/m·K. By comparison, copper has only 385 W/m·K.

Nanotubes are produced about 60 tons/years now (2007). Price is about $100 - 50,000/kg. Experts predict production of nanotubes 6000 tons/years and price $1 – 100/kg to 2012.

The artificial fibers are cheap and widely used in tires and everywhere. The author has found only old information about textile fiber for inflatable structures (Harris J.T., Advanced Material and Assembly Methods for Inflatable Structures, AIAA, Paper No. 73-448, 1973). This refers to DuPont textile Fiber **B** and Fiber **PRD-49** for tire cord. They are 6 times strong as steel (psi is 400,000 or 312 kg/mm$^2$) with a specific gravity only 1.5. Minimum available yarn size (denier) is 200, tensile module is $8.8 \times 10^6$ (**B**) and $20 \times 10^6$ (**PRD-49**), and ultimate elongation (percent) is 4 (**B**) and 1.9 (**PRD-49**). Some data are in Table 1.

**Table 1.** Material properties

| Material / Whiskers | Tensile strength kg/mm$^2$ | Density g/cm$^3$ | Fibers | Tensile strength kg/mm$^2$ | Density g/cm$^3$ |
|---|---|---|---|---|---|
| AlB$_{12}$ | 2650 | 2.6 | QC-8805 | 620 | 1.95 |
| B | 2500 | 2.3 | TM9 | 600 | 1.79 |
| B$_4$C | 2800 | 2.5 | Allien 1 | 580 | 1.56 |
| TiB$_2$ | 3370 | 4.5 | Allien 2 | 300 | 0.97 |
| SiC | 1380-4140 | 3.22 | Kevlar or Twaron | 362 | 1.44 |
| **Material** | | | Dynecta or Spectra | 230-350 | 0.97 |
| Steel pre-stressing strands | 186 | 7.8 | Vectran | 283-334 | 0.97 |
| Steel Piano wire | 220-248 | | E-Class | 347 | 2.57 |
| Steel A514 | 76 | 7.8 | S-Class | 471 | 2.48 |
| Aluminum alloy | 45.5 | 2.7 | Basalt fiber | 484 | 2.7 |
| Titanium alloy | 90 | 4.51 | Carbon fiber | 565 | 1,75 |
| Polypropylene | 2-8 | 0.91 | Carbon nanotubes | 6200 | 1.34 |

Source: [16]-[19]. Howatsom, A.N., Engineering Tables and Data, p.41.

Industrial fibers have $\sigma$ = 500-600 kg/mm$^2$, $\gamma$ = 1800 kg/m$^3$, and $\sigma/\gamma$ = 2,78x10$^6$. But we are used in all our projects the cheapest films and cables (safety $\sigma$ = 50 - 100 kg/mm$^2$).

**5. Wind effect.** As wind flows over and around a fully exposed, nearly completely sealed inflated dome, the weather affecting the external film on the windward side must endure positive air pressures as the wind stagnates. Simultaneously, low air pressure eddies will be present on the leeward side of the dome. In other words, air pressure gradients caused by air density differences on different parts of the dome's envelope is characterized as the "buoyancy effect". The buoyancy effect will be greatest during the coldest weather when the dome is heated and the temperature difference between its interior and exterior are greatest. In extremely cold climates such as the Arctic and Antarctic Regions the buoyancy effect tends to dominate dome pressurization.

**6. Solar radiation.** Solar radiation impinging the orbiting Earth is approximately 1400 W/m$^2$. The average Earth reflection by clouds and the sub-aerial surfaces (water, ice and land) is about 0.3. The Earth-atmosphere absorbs about 0.2 of the Sun's radiation. That means about $q_0$ = 700 W/m$^2$s of solar energy (heat) reaches our planet's surface in cloudy weather at the Equator. That means we can absorb about 30 - 80% of solar energy. It is enough for normal plant growth in wintertime (up to 40-50° latitude) and in circumpolar regions with a special variant of the dome design.

The solar spectrum is graphically portrayed in Fig. 5.

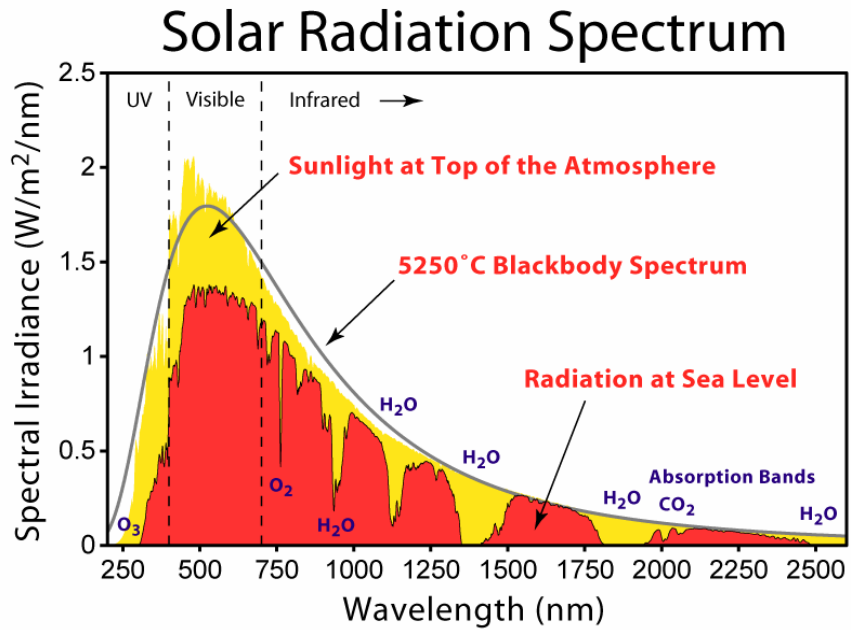

**Fig. 5**. Spectrum of solar irradiance outside atmosphere and at sea level with absorption of electromagnetic waves by atmospheric gases. Visible light is 0.4 - 0.8 µ (400 – 800 nm).

The visible part of the Sun's spectrum is only $\lambda$ = 400 – 800 nm (0.4 to 0.8 $\mu$.). Any warm body emits radiation. The emission wavelength depends on the body's temperature. The wavelength of the maximum intensity (see Fig. 5) is governed by the black-body law originated by Max Planck (1858-1947):

$$\lambda_m = \frac{2.9}{T}, \quad [mm], \qquad (1)$$

where $T$ is body temperature, °K. For example, if a body has an ideal temperature 20 °C ($T$ = 293 °K), the wavelength is $\lambda_m$ = 9.9 µ.

The energy emitted by a body may be computed by employment of the Josef Stefan-Ludwig Boltzmann law:

$$E = \varepsilon \sigma_s T^4, \quad [W/m^2], \qquad (2)$$

where $\varepsilon$ is coefficient of body blackness ($\varepsilon = 0.03 \div 0.99$ for real bodies), $\sigma_s = 5.67 \times 10^{-8}$ [W/m$^2$·K] Stefan-Boltzmann constant. For *example*, the absolute black-body ($\varepsilon = 1$) emits (at $T = 293\ ^0$K) the energy $E = 418$ W/m$^2$.

**7. Earth's atmosphere**. The property of Earth's atmosphere needed for computations are presented in Table 2 below.

**Table 2.** Standard Earth atmosphere

| $H$ km | 0 | 1 | 2 | 3 | 4 | 5 | 6 | 7 |
|---|---|---|---|---|---|---|---|---|
| $\bar{p} = p_h/p_o$ | 1 | 0.887 | 0.784 | 0.692 | 0.609 | 0.533 | 0.466 | 0.406 |
| $H$ km | 8 | 9 | 10 | 11 | 12 | 13 | 14 | 15 |
| $\bar{p} = p_h/p_o$ | 0.362 | 0.304 | 0.261 | 0.224 | 0.191 | 0.164 | 0.14 | 0.12 |

## b) General theory

**1. Effect of nuclear explosion.** Effect of nuclear explosion depends from many factors. For example, from atmosphere conditions. The results of computation for typical conditions [15] are presented in Table 3. CONVERSION NOTE: 1 psi = 0.45359 kilogram per square inch.

**Table 3**. Effects of nuclear explosion

| Yield (megatons) | 0.001 | 0.01 | 0.1 | 1 | 10 |
|---|---|---|---|---|---|
| Thermal radiation radius (3$^{rd}$ degree burns) | 687 m | 1.8 km | 4.5 km | 11.7 km | 30 km |
| Air blast radius (4.6 psi, widespread destruction) | 739 m | 1.6 km | 3.4 km | 7.2 km | 15.4 km |
| Air blast radius (20 psi, near-total fatalities) | 280 m | 599 m | 1.3 km | 2.7 km | 5.9 km |
| Ionizing radiation radius (500 rem) | 840 m | 1.3 km | 2 km | 3.1 km | 4.8 km |
| Fireball duration | 0.2 s | 0.6 s | 1,6 s | 4.5 s | 12.7 s |
| Fireball radius (minimum) | 30 m | 70 m | 170 m | 430 m | 1.1 km |
| Fireball radius (airburst) | 30 m | 80 m | 210 m | 530 m | 1.3 km |
| Fireball radius (ground-contact airburst) | 40 m | 110 m | 280 m | 700 m | 1.8 km |

**Notes to Table 3:**

- All figures assume optimum burst height.
- Thermal radiation is non-ionizing electromagnetic radiation which has a significant heating effect. Air is virtually transparent to thermal radiation. At the destructive radius, the thermal radiation intensity is sufficient to cause lethal burns.
- The first air blast is 4.6psi overpressure, which is sufficient to collapse most residential and industrial structures. Note that exposed humans can actually survive such a blast, about 1/3 bar above standard. However, that much pressure exerted against the face of a building exerts very high force (a 40 foot tall, 50 foot wide structure would be hit with more than 600 tons-force).
- The second air blast category is 20 psi over-pressure, which is sufficient to destroy virtually any large above-ground structure and cause nearly 100% fatalities.
- Ionizing radiation is electromagnetic radiation of sufficient frequency (and hence energy) to literally "knock off" electrons from atoms, thus ionizing them. Ionizing radiation is extremely dangerous but it is also strongly absorbed by air, unlike thermal radiation. At the 500rem dosage, mortality is between 50% and 90%, although this can be mitigated with prompt and sophisticated medical care (which may not be available in the aftermath of a nuclear attack).

- Fireball duration is based on emission intensity reduction to 10% of peak.

- Fireball radius is based on a scaling law from "The Effects of Nuclear Weapons" (1977), Chapter IIc, from excerpts reprinted at EnviroWeb. According to that source, fireball radius scales with (Y^0.4), where Y is yield. Also note that a ground-contact airburst creates a larger fireball because some of the energy is reflected back up from the surface.

The data used in HYDESim are based on information found in "The Effects of Nuclear Weapons", 3rd Edition, by Samuel Glasstone and Philip J. Dolan [15].

### Over-pressure Key

- **15 psi** Complete destruction of reinforced concrete structures, such as skyscrapers, will occur within this ring. Between 7 psi and 15 psi, there will be severe to total damage to these types of structures.
- **5 psi** Complete destruction of ordinary houses, and moderate to severe damage to reinforced concrete structures, will occur within this ring.
- **2 psi** Severe damage to ordinary houses, and light to moderate damage to reinforced concrete structures, will occur within this ring.
- **1 psi** Light damage to all structures, and light to moderate damage to ordinary houses, will occur within this ring.
- **0.25 psi** Most glass surfaces, such as windows, will shatter within this ring, some with enough force to cause injury.

Results of air blast and the thermal radiation computation via bomb yield are presented in Fig. 6.

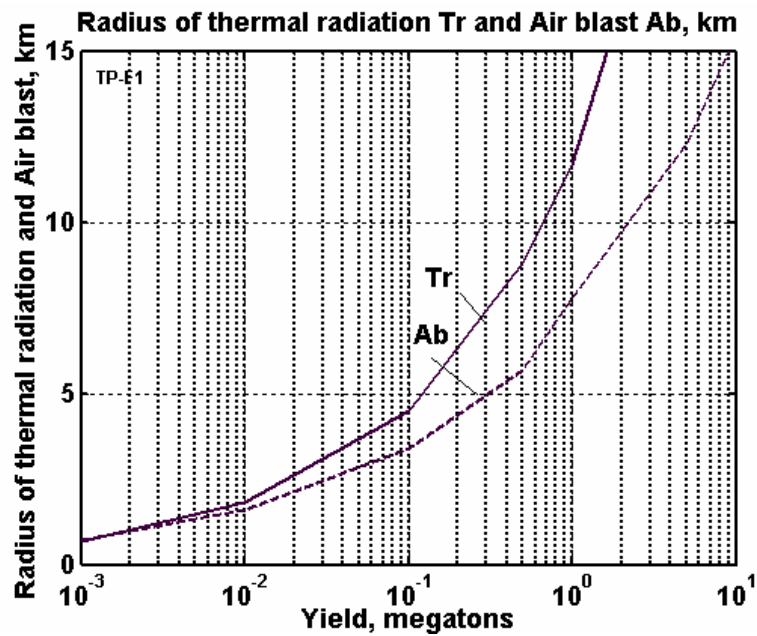

**Fig.6.** Thermal radiation radius (3$^{rd}$ degree burns) and Air blast radius (4.6 psi, wide-spread destruction) via bomb yield.
Reminder: the 1945 Hiroshima bomb had yield about 0.015 Mt, the hydrogen bomb has 0.1 (and more) Mt.

**2. Decrease of bomb effects with increased detonation altitude.** The decreasing of bomb efficiency from height of explosion may be estimated by equations:

$$T_r = \left(\frac{R}{H}\right)^3, \quad A_b = \left(\frac{p_h}{p_0}\right)\left(\frac{A}{H}\right)^3, \tag{3}$$

where $T_r$ is relative thermal radiation; $R$ is thermal radiation (3$^{rd}$ degree burns), km; $A_b$ is Air blast radius (4.6psi, widespread destruction), km; $H$ is altitude of explosion, km.
Result of computation is presented in Fig.7.

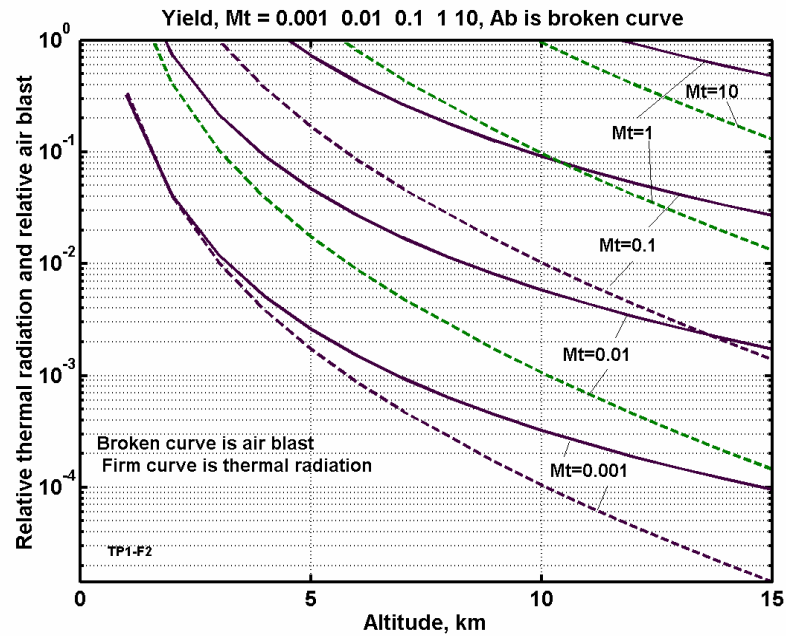

**Fig.7.** Decreasing of bomb effect versus altitude for different bomb yield (Mt is Megatons). The broken curve is the thermal effect, the firm line is the air blast effect.

As you see the 10 kt bomb exploded at altitude 10 km decreases the air blast effect about in 1000 times and thermal radiation effect without the second cover film in 500 times, with the second reflected film about 5000 times. The hydrogen 100kt bomb exploded at altitude 10 km decreases the air blast effect about in 10 times and thermal radiation effect without the second cover film in 20 times, with the second reflected film about 200 times.

Only power 1000kt thermonuclear (hydrogen) bomb can damage city. But this damage will be in 10 times less from air blast and in 10 times less from thermal radiation. If the film located at altitude 15 km, the damage will be in 85 times less from the air blast and in 65 times less from the thermal radiation.

For protection from super thermonuclear (hydrogen) bomb we need in higher dome altitudes (20-30 km and more). We can cover by AB-Dome the important large region and full country.

**3. Warhead overload when film break.** When the impacting warhead penetrates the city-shielding film, the warhead breaks the film. The film break is made at a very high speed (3-6 km/s) and, therefore, the penetrating warhead has instantaneous overload. This overload appears from the film tensile stress and from part of the film which get instantaneous acceleration.

The overload from film break may be estimated by equation

$$a = F/gM, \quad F \approx \sigma_{max} \delta L, \tag{4}$$

where $a$ is overload in $g$; $g = 9.81$ m/s$^2$ is Earth's acceleration; $M$ is warhead mass, kg; $F$ is break force, N; $\sigma_{max}$ is maximum tensile stress of film, N/m$^2$; $\delta$ is film thickness, m; $L$ is length of break, m.

For *example*, if the warhead makes a single hole with only a 1 m diameter ($L \approx 3$ m), $\sigma_{max} = 250$ kg/mm$^2 = 2.5 \times 10^9$ N/m$^2$, $\delta = 0.25$ mm $= 0.00025$ m, Warhead mass $M = 100$ kg. We receive the $F = 19 \times 10^5$ N $=190$ metric tons, $a = 1900$ g. That is a gigantic material overload! Only special (high mass) design of warhead can possibly successfully resist this instantaneous pressure overload. The film mass $m \approx 0.4$ kg/m$^2$, adding an additional destructive overload to the incoming warhead (as 1 kg of explosive!).

Who doubt, one can take conventional customer plastic shopping bag, roll it into a tube shape and break it. Note, the polyethylene bag has thickness only $\delta = 7$ -12 μm, $L \approx 0.5$ and not a high $\sigma$.

The film produces a dispersed stress. If the film is armored by strong stones, the 1 – 5 stones (in example over) full destroyed the any bomb, missile, rocket, and aircraft because the relative speed of flight apparatus is closed to artillery projectiles. In high-speed blow, the mass is a more important factor than the strength of stone studding the AB-Dome. The author saw the result of blow test when a small (some grams) stone struck with speed 7 km/s into an alumina armor of 10 cm thickness. One made a hole of 10 cm diameter!
The results of computation of equations (4) are shown in Fig.8.

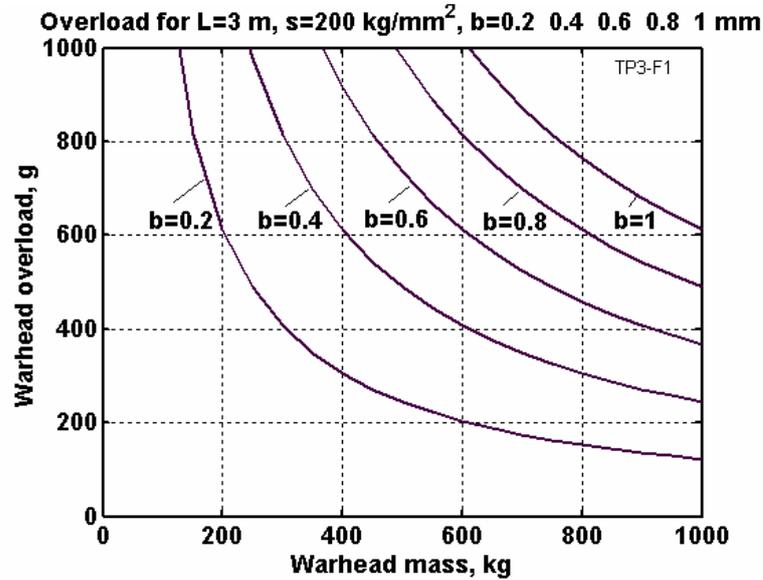

**Fig.8**. Warhead percussion overload when warhead strike in cover dome film via warhead mass for different the film thickness b = $\delta$, maximal film tensile stress s = $\sigma$ = 200 kg/mm$^2$, and length of rupture $L$ = 3 m.

**4. The thickness and weight of the AB-Dome**, its sheltering shell of film, is computed by formulas (from equation for tensile strength):

$$\delta_1 = \frac{Rp}{2\sigma}, \quad \delta_2 = \frac{Rp}{\sigma}, \tag{5}$$

where $\delta_1$ is the film thickness for a spherical dome, m; $\delta_2$ is the film thickness for a cylindrical dome, m; $R$ is radius of dome or radius of cover cell between cable (it may be half of distance between top cable), m; $p$ is additional pressure into the dome, N/m$^2$, ($p$ depends from altitude); $\sigma$ is safety tensile stress of film, N/m$^2$.

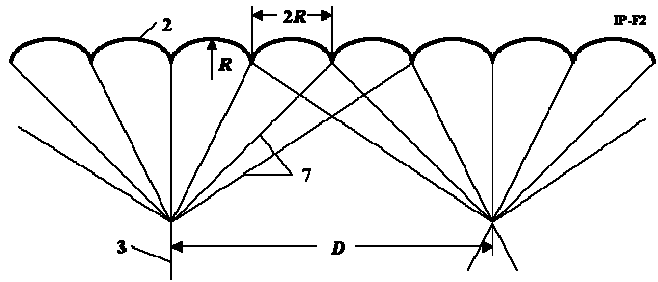

**Fig. 9**. Cable-retaining system. Radius, $R$, spherical cell of dome cover and distance $D$ between main cable.

For *example*, compute the film thickness for dome having radius $R$ = 100 m (distance between top cable 7 is 400 m), additional air pressure $p$ = 0.01 atm ($p$ = 1000 N/m$^2$), safety tensile stress $\sigma$ = 50 kg/mm$^2$ ($\sigma$ = 5×10$^8$ N/m$^2$), hemi-spherical dome. We receive $\delta_1$ = 0.1 mm. Distance between main cable 3 is $D$ = 0.8 km (Fig.9).
The computation for others case are presented in Fig. 10 below.

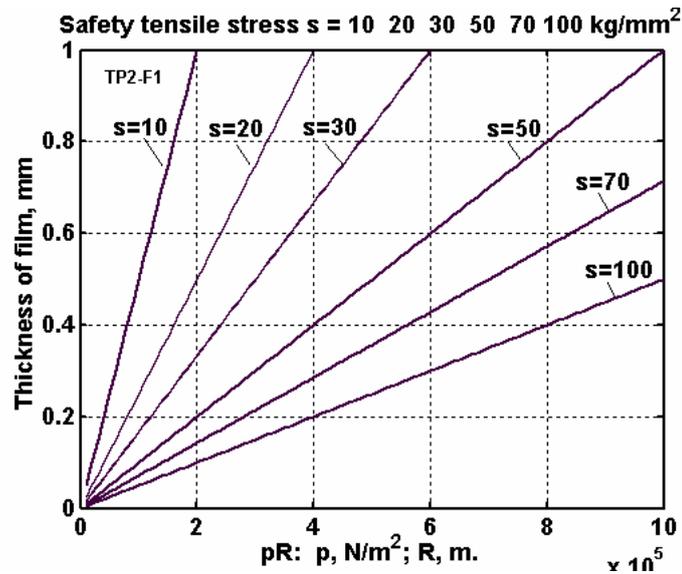

**Fig.10.** The thickness of top cover via the production of overpressure and radius spherical dome cell (distance between top cables for different safety film tensile stress.

The cover weight (mass) of 1 m² is computed by equation

$$m = \gamma \delta, \qquad (6)$$

where $m$ is 1 m² film mass, kg/m²; $\gamma$ is cover density, m. For *example*, if the cover thickness is $\delta = 0.2$ mm $= 0.0002$ m and $\gamma = 1500$ kg/m³, the $m = 0.3$ kg/m².
Area $S_c$ of semi-sphere diameter $R$, film cover mass $M_f$ and cost $C$ of AB-Dome plastic envelope are

$$S_c = 2\pi R^2, \quad M_f = m_f S_c, \quad C = c S_c, \quad C = c_m M_f, \qquad (7)$$

where $R$ is radius of hemisphere, m; $m_f$ is average cover area of 1 m²; $c$ is cost of 1 m², $US/m²; $c_m$ is cover cost of 1 kg, $US/m²; $C$ is cost of total cover, $US.

*Example*. Let us take the semi-sphere Dome. If $m_f = 0.3$ kg/m², film cost $c = \$0.1$ /m². The film mass covered of 1 km² of ground area is $M_1 = 2\times 10^6$ $m_c = 600$ tons/km² and film cost is $60,000/km².

The area of big city diameter 20 km is 314 km². Area of semi-spherical dome is 628 km². The cost of Dome cover is 62.8 millions $US. We can take less the overpressure ($p = 0.001$ atm) and decrease the cover cost in 5 – 7 times.

The total cost of installation is about 30-90 million $US.

That is less in hundred times, than the cost of anti-rocket system (tens of billions $U.S.A.). The anti-rocket system is useless in peacetime and it may be useless soon in wartime because the weapon is permanently improved. The offered AB-Dome is very useful in peacetime (control weather, temperature inside!), The AB-Dome defense also against any biological, chemical, radioactivity-emitting weapons. AB-Dome is CLOSED-LOOP system. All Earth can be poisoned by radioactive precipitations, poison-gas, harmful microbes but a big city (country) can exist under AB-Dome. (That is the idea and use of the "Doomsday Weapon".)

**5. The mass of holding cable for 1 m² projection of AB-Dome**. The mass of the support cable for every projection 1 m² of dome cover may be computed by equation:

$$m_c \approx \gamma_c \frac{\overline{pp}}{\sigma} H, \quad M_c = \sum_S m_c, \quad M_c \approx \gamma_c \frac{p_a}{\sigma} S H_a, \qquad (8)$$

where $m_c$ is cable mass supported the projection one m² of the dome cover, kg/m²; $M_c$ is total mass of cable, kg; $S = \pi R^2$ is projection of cover on ground, m²; $\gamma_c$ is density of cable, kg/m³; $p$ is overpressure, N/m²; $\sigma$ is safety cable tensile stress, N/m²; $H$ is height of cable, m; $\bar{p}$ is relative air pressure at given altitude (Table 2); $p_a$ is average over pressure, N/m²; $H_a$ is average height of the support cable, m.

*Example*, for cable having the $\sigma = 3 \times 10^9$ N/m², $\gamma_c = 1500$ kg/m³ (Table 1), $H = 10$ km $= 10^4$ m, ($\bar{p} = 0.261$), $p = 0.01$ atm $= 1000$ N/m²; the mass of cable is $m_c = 1.3$ kg/m². If we take $p = 0.001$ atm $= 100$ N/m², $R = 10$ km, $H_a = 5$ km, average overpressure will be $p_a = 63$ N/m², $\sigma = 10^9$ N/m², than the mass of support cable is about 1500 tons.

*We can design the dome cover without the support cable*. In this case we compute the thickness of dome cover for radius $R$ of full Dome (see Eq.(5)). The cover thickness will be more. That is better for defense from warhead. If we wand compute more exactly and spend less cover mass, we must compute the variable overload (from altitude). In this case we received the variable thickness of dome cover.

### 6. Lift force of 1 m² projection of dome cover.
The lift force of 1 m² projection of dome cover is computed by equation:

$$L_1 = \bar{p} p , \qquad (9)$$

where $L_1$ is lift force of 1 m² projection of dome cover, N/m²; $p$ is overpressure, N/m²; $\bar{p}$ is relative air pressure at given altitude (Table 2).

*Example*, for $p = 0.01$ atm $= 1000$ N/m², $H = 10$ km ($\bar{p} = 0.261$, Table 2) we get $L_1 = 261$ N $= 26$ kg.

Result of computation for different $p$ is shown in Fig.11.

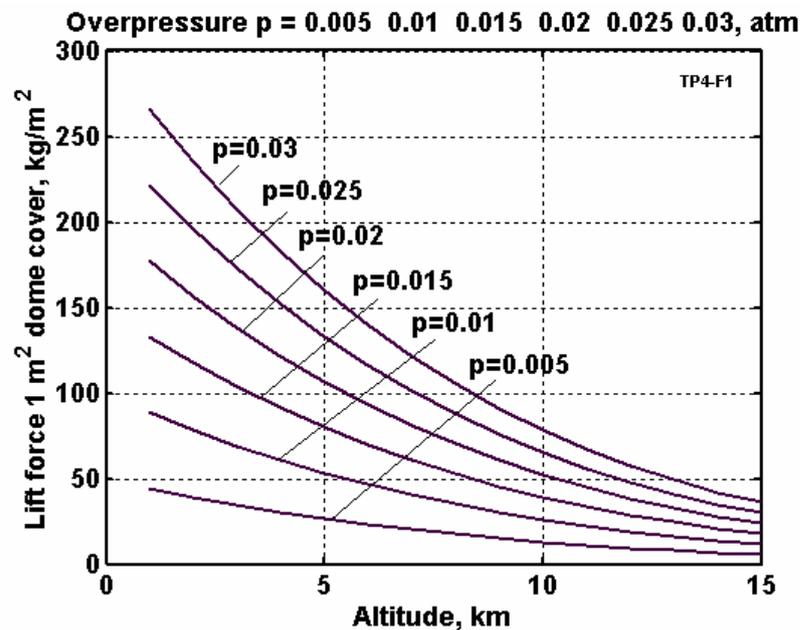

**Fig.11**. Lift force 1 m² of vertical projection the AB-Dome versus altitude for different over-pressures.

1. **Leakage of air through a hole.** The leakage of air through hole, requested power of ventilator, and time of sinking of Dome cover (in case large of hole) may be estimates by equation:

$$V = \sqrt{\frac{2p}{\rho}}, \quad M_a = \rho V S_h, \quad N = \frac{p V S_h}{\eta}, \qquad (10)$$

where $V$ is speed of air leakage, m/s; $p$ is overpressure, N/m²; $\rho$ is air density at given altitude, $\rho = 1{,}225$ kg/m³ at $H = 0$; $S_h$ is area of hole, m; $N$ is motor power, W; $\eta$ is coefficient efficiency of motor.

*Example*. The area of hole equals $S_h = 10^2$ m$^2$ (10×10 m) at $H = 0$ m, $p = 0.001$ atm = 100 N/m$^2$, $\eta = 0.8$. Computation gives the $V = 12.8$ m/s, $M_a = 1568$ kg/s, $N = 161$ kW.

Let us to estimate the time of Dome cover sinking if no supercharge (pumping). Take the sphere radius $R = 500$ m. The volume of semi-sphere is $v = 262 \times 10^6$ m$^3$, the air rate is $q = VS_h$ = 12.8×100 = 1280 m$^3$. The time of the full sinking is $t = v/q = 204 \cdot 10^3$ s = 57 hours.

Note, the overpressure air will be flow out from dome. That means the radioactive bomb pollution cannot penetrate to dome. The outer air pumping in dome can be filtered from radioactive dust.

The repair of dome hole is easy. The support cable of the need part of dome is reeled on and film patch closes the hole.

**9. Protection studding stones.** For increasing protection the AB-Dome cover may be armored with selected strong stones (for example, pebbles, rock flakes, pieces of concrete, and other cheap and available materials), having mass 0.1 – 1 kg. The current anti-rocket systems widely used the kinetic method for destroying the strategic missiles and warhead. The anti-rocket scatters the small balls in the trajectory of rocket (warhead) and when the rocket (warhead) strike ball with very high speed (a lot of times more than conventional projectile, bullet, shell, explpsove) the missile is destroyed. It is not necessary to apply explosive.

The number of the stones for 1 km$^2$ square area may be estimated by following way. Assume the step of stones is 1 m. Then the number of stones in 1 km$^2$ is $N_s \approx L^2$, where length $L = 1000$ m is side of square. If the average mass of a stone is 0.2 kg, the total mass of stones in 1 km$^2$ will be 200 tonnes per square kilometer. If we then put an additional, similar stone net in this same region, then the stone mass increases by two times and will be 400 tons/km$^2$, but stone step will be equal about 0.5 m. That is more than enough for destroying an incoming warhead. If warhead has cross-section area 1 m$^2$ then 1-5 stones will destroy it.

**10. The wind dynamic pressure** is:

$$p_w = \frac{\rho V^2}{2}, \qquad (11)$$

where $\rho = 1.225$ kg/m$^3$ is air density; $V$ is wind speed, m/s.

For example, a storm wind with speed $V = 20$ m/s, standard air density is $\rho = 1.225$ kg/m$^3$. Than dynamic pressure is $p_w = 245$ N/m$^2$. That is four time less when internal pressure $p = 1000$ N/m$^2$ = 0.01 atm. When the need arises, sometimes the internal pressure can be voluntarily decreased, bled off.

## Macro-project

Let us to consider the protection of typical big city having diameter 20 km by semi-spherical AB-Dome, having radius (altitude) $R = 10$ km. The covered region is $S = \pi R^2 = 314$ km$^2$, the semi-spherical area is $S_s = 2S = 628$ km$^2$. The world's biggest city (Mumbai, India) has an area of 484 km$^2$ and a population of about 14.3 million persons. One can be protected by spherical AB-Dome having radius $R = 12 – 15$ km.

The many computations for this Dome are made as examples in theoretical section above. We summarize these data in one macro-project below.

Let us take the overpressure 0.01 atm at sea level. Note, a man (deep diver) can keep overpressure some atmosphere. The Earth's atmosphere changes the pressure in some percents and we do not feel it. Take a film having safety tensile stress $\sigma = 50$ kg/mm$^2$, local $R = 100$ m (distance between main cable is 0.8 km, Fig. 9), then the film has a thickness $\delta_1 = 0.1$ mm at sea level $H = 0$ and $\delta_1 = 0.026$ mm at altitude $H = 10$ km [Eq. (5)]. The average $\delta_1 = 0.063$ mm.

Area of semi-sphere is $S_s = 2\pi R^2 = 628$ km². If density of film equals $\gamma = 1500$ kg/m³, the total mass of cover film is $M = \gamma \delta S_s = 60,000$ tons. If cost of film is $1/kg, the total cost of top cover is C = $60 millions. Take the thickness of lower film $\delta = 0.01$ mm and density $\gamma = 1500$ kg/m³, then the mass of lower film is $M = \gamma \delta S = 4,700$ tons and cost $4.7 millions.

If the stone step is $s = 0.5$ m and stone mass equals $m_s = 0.2$ kg each, we early calculated 400 tons/km² of stones, the needed total mass of stone is $M_s = m_s S_s = 251,200$ tons. If stones price is U.S.A. $10/ton, the stone costs $2.5 millions.

The average overpressure is $p = 0.0063$ atm = 630 N/m². The total cover lift force is $L = pS = 20$ millions tons. If support cables has safety tensile stress $\sigma = 100$ kg/mm², $\gamma = 1500$ kg/m³, price $0.5/kg and average height 5 km, the total mass of support cable is $M_c = 150,000$ tons, the cost of cables is $75 millions.

The total mass of construction is 466,000 tons and cost of construction material $142 millions. The total cost of AB-Dome cover big city is about $170 millions.

Our dome lift force (20,000,000 metric tons) in 43 times is over then a total weight of the dome construction (466,000 tonnes). If we decrease the overpressure in 10 times ($p = 0.001$ atm), the need mass of construction material decreases approximately in 10 times. The AB-Dome will be cost about $30 millions.

As you see the most mass and cost have the support cables. If we use the more thickness or strong film ($\sigma = 100$ kg/mm²) we can make the AB-Dome without support cable. Let us to make estimation of this case for overpressure $p = 0.001$ atm. The average thickness of top film is 0.63 mm, the cover mass is 600,000 tonnes. The cost of construction material is $75 millions (for the cost of top cover $0.5/kg). The total cost of Dome is about $90 millions. The total dome lift force is about $L = 2,000,000$ tons. The non-cable Dome is more expensive about in 3 times (in comparison $30 millions). However, the AB-Dome without internal cables is more comfortable for internal helicopter flight, but one can be less comfortable for repair. The non-cable AB-Dome requests a more cover thickness (by about 3-4 times), which increases the overload of the warhead and may make the stones unneeded.

Our Dome is far from optimal. The average lift force (6.3 kg/m²) is over in 4.7 times the weight of cover (0.94 kg/m²) plus weight of stones (o.4 kg/m²). We can increase the mass (and number) of stones in 3-5 times.

The author is prepared to discuss the problems with organizations which are interested in research and development related projects.

## Discussion

As you see the 10 kt bomb exploded at altitude 10 km decreases the air blast effect about in 1000 times and thermal radiation effect without the second cover film in 500 times, with the second reflected film about 5000 times. The hydrogen 100kt bomb exploded at altitude 10 km decreases the air blast effect about in 100 times and thermal radiation effect without the second cover film in 10 times, with the second reflected film about 100 times.

Only power 1000kt (1 Megaton) thermonuclear (hydrogen) bomb can damage a city. But this damage will be in 10 times less from air blast and in 10 times less from thermal radiation. If the film located at altitude 15 km, the damage will be in 85 times less from the air blast and in 65 times less from the thermal radiation. The white clouds (they are located at altitude 2-4 km inside the AB-Dome) additionally decreases the thermal radiation effect. For more security the cover must be located at altitude up 20 km .

When warhead penetrates the film, the warhead breaks the film. The break is made at extremely high speed (3-6 km/s) and warhead has instantaneous overload. This over-load appears from the film tensile stress and from part of the film which get instantaneous acceleration. The over-load reaches 1000g (and more) and destroys any warhead, missile, rocket, and aircraft. The stones have energy (relative warhead) in 2 – 3 times more them explosive of same mass.

The area of a big city with a diameter of 20 km is about 314 km$^2$. The total cost of the AB-Dome installation is about 30-90 million $U.S.A.

That is less by hundred times, then the cost of anti-rocket system (tens of billions $U.S.A.). The anti-rocket system is useless in peacetime and it may be useless soon in wartime because the weapon is permanently improved. The offered AB-Dome is very useful in peacetime because it can be used to control climate, weather and the air temperature inside. The AB-Dome defense also works against any biological, chemical, radioactive outer weapons. AB-Dome is CLOSED-LOOP system. All Earth can be poisoned by radioactive precipitations, poison-gas, harmful microbe, but a big city (country) can exist under AB-Dome.

Note, the over-pressure air will flow out from dome. That means the radioactive bomb pollution cannot penetrate to dome in case of AB-Dome (cover) damage. The expelling air pumping in dome can be filtered from radioactive dust.

Constructing the AB-Dome is rather easy, as building macro-projects go. The covering spreads on the ground, air pump turned on and the air-supported AB-Dome then rises.

The repair of an AB-Dome hole is not difficult. The support cable of the need part of dome is reeled on and film patch closes the hole. If the dome does not have the support cable, one can have suspended cables which allow reaching any part of AB-Dome. The non-cable AB-Dome requests a more cover thickness (about in 3-4 times), which increases the overload of the warhead and may make the stones unneeded.

For increasing protection, the dome cover may be armored with many strong stones (for example, pebbles, rock flakes, pieces of concrete, and other selected cheap materials), having mass 0.1 – 1 kg. The current anti-rocket systems widely used the kinetic method for destroying the strategic missiles and warhead. The anti-rocket scatters the small balls in the trajectory of rocket (warhead) and when the rocket (warhead) strikes a ball with very high speed (a lot of times more then conventional projectile, bullet, shell) the missile is destroyed. It is not necessary to apply explosive.

The control of Earth's regional and global weather is an important problem for humanity. That ability dramatically increases the territory suitable for people to live in, the sown area, and crop capacity. In the long term, it allows us to convert all Earth land such as Alaska, North Canada, Siberia and deserts such as the Sahara or the Gobi into a prosperous garden. The suggested method is very cheap (cost of covering 1 m$^2$ is about 2 - 15 U.S.A. cents) and may be utilized at the present time. We can start from a small area, from small towns in bad climactic regions and extended to a large area.

Film domes can foster the fuller economic development of cold regions such as the Earth's Arctic and Antarctic and, thus, increase the effective area of territory dominated by humans. Normal human health can be maintained by ingestion of locally grown fresh vegetables and healthful "outdoor" exercise. The domes can also be used in the Tropics and Temperate Zone. Eventually, they may find application on the Moon or Mars since a vertical variant, inflatable towers to outer space, are soon to become available for launching spacecraft inexpensively into Earth-orbit or interplanetary flights. AB-Dome can keep at high altitude the load up 300 kg/sq. m.

Lest it be objected that such AB-Domes would take impractical amounts of plastic, consider that the world's plastic production is today on the order of 100 million tons. If, with economic growth, this amount doubles over the next generation and the increase is used for doming over territory, at 300-500 tons a square kilometer 200,000 square kilometers could be roofed over annually. While small in comparison to the approximately 150 million square kilometers of land area, consider that 200,000 one kilometer sites scattered over the face of the Earth newly made productive and more habitable could revitalize vast swaths of land surrounding them—one square kilometer here could grow local vegetables for a city in the desert, one over there could grow biofuel, enabling a desolate South Atlantic island to become fuel independent; at first, easily a billion people a year could be taken out of sweltering heat, biting cold and slashing rains, saving the money buying and running heating and air conditioning equipment would require. Additionally, clean rain water could flow directly to cisterns, away from the pollution of the storm sewers.

In effect, by doming over inhospitable land as specified, in exchange we get new territory for living with a wonderful climate.

The associated problems are researched in references [1]-[12].

## Results

Author offers the cheap AB-Dome which protects the big cities from nuclear, chemical, biological weapon (bombs) delivered by warheads, strategic missiles, rockets, and aviation. The offered AB-Dome is also very useful in peacetime because that protests the city from outer weather and creates a fine climate into Dome.

Main advantages of offered AB-Dome:
1. AB-Dome is cheaper in hundreds times then current anti-rocket systems.
2. AB-Dome does not need in high technology and can build by poor country.
3. It is easy for building.
4. Dome is used in peacetime; it creates the fine climate (weather) into Dome.
5. AB-Dome protects from nuclear, chemical, biological weapon.
6. Dome produces the autonomous existence of the city population after total World nuclear war and total confinement (infection) all planet and its atmosphere.
7. Dome may be used for high region TV, for communication, for long distance locator, for astronomy (telescope).
8. Dome may be used for high altitude tourism.
9. Dome may be used for the high altitude windmills (getting of cheap renewable wind energy).
10. Dome may be used for a night illumination and entertainment.

The additional applications of offered AB-Dome the reader finds in [1]-[12].

## *Acknowledgement*

The author wishes to acknowledge Richard B.Cathcart for correcting the English and offering useful advice.